\documentclass[reprint,aps,prb,amsmath,superscriptaddress,secnumarabic,amssymb, nobibnotes]{revtex4-1}
\usepackage{graphicx}

\begin{document}

\title{Absorption and generation of femtosecond laser-pulse excited spin currents in non-collinear magnetic bilayers}

\author{M.L.M. Lalieu}
\email[Corresponding author: ]{m.l.m.lalieu@tue.nl}
\affiliation{Department of Applied Physics, Institute for Photonic Integration, Eindhoven University of Technology, P.O. Box 513, 5600 MB Eindhoven, The Netherlands}
\author{P.L.J. Helgers}
\affiliation{Department of Applied Physics, Institute for Photonic Integration, Eindhoven University of Technology, P.O. Box 513, 5600 MB Eindhoven, The Netherlands}
\author{B. Koopmans}
\affiliation{Department of Applied Physics, Institute for Photonic Integration, Eindhoven University of Technology, P.O. Box 513, 5600 MB Eindhoven, The Netherlands}
\date{\today}
\begin{abstract}
Spin currents can be generated on an ultrafast timescale by excitation of a ferromagnetic (FM) thin film with a femtosecond laser-pulse. Recently, it has been demonstrated that these ultrafast spin currents can transport angular momentum to neighbouring FM layers, being able to change both the magnitude and orientation of the magnetization in the adjacent layer. In this work, both the generation and absorption of these optically excited spin currents are investigated. This is done using non-collinear magnetic bilayers, i.e.\ two FM layers separated by a conductive spacer. Spin currents are generated in a Co/Ni multilayer with out-of-plane (OOP) anisotropy, and absorbed by a Co layer with an in-plane (IP) anisotropy. This behaviour is confirmed by careful analysis of the laser-pulse induced magnetization dynamics, whereafter it is demonstrated that the transverse spin current is absorbed very locally near the injection interface of the IP layer ($90\%$ within the first $\approx 2$ nm). Moreover, it will also be shown that this local absorption results in the excitation of THz standing spin waves within the IP layer. The dispersion measured for these high frequency spin waves shows a discrepancy with respect to the theoretical predictions, for which a first explanation involving intermixed interface regions is proposed. Lastly, the spin current generation is investigated using different number of repeats for the Co/Ni multilayer, which proves to be of great relevance for identifying the optical spin current generation mechanism.
\end{abstract}
\maketitle

\section{Introduction}
The discovery of ultrafast demagnetization in ferromagnetic (FM) thin films after femtosecond (fs) laser-pulse excitation induced a growing interest in the field of fs magnetization dynamics. Two decades ago Beaurepaire et al. demonstrated that the magnetization in a Ni thin film can be quenched to almost half its initial value within a picosecond after the fs laser-pulse excitation \cite{Beaurepaire1996}. Although the discovery triggered and interesting debate on the physical mechanism responsible for the rapid loss of magnetization \cite{Koopmans2009,Mueller2013,Carva2013,Battiato2010,Battiato2012}, its relevance for fast and energy efficient magnetic data storage was quickly recognized. This realization eventually lead to the discovery of all-optical magnetization reversal in both (synthetic) ferrimagnets \cite{Stanciu2007,Mangin2014} and ferromagnets \cite{Lambert2014}.

A third important discovery was that spin currents are generated upon excitation of a FM film. This was first demonstrated in a collinear magnetic bilayer, where angular momentum transfer through the spacer layer resulted in a faster and larger demagnetization of the two anti-parallel FM layers \cite{Malinowski2008}. Several more recent studies have confirmed these laser-pulse induced spin currents \cite{Melnikov2011,Rudolf2012,Alekhin2014,Schellekens2014,Choi2014,Choi2015}. It even has been claimed that the optically excited spin current can enhance the magnetization in one of the FM layers of the magnetic bilayer \cite{Rudolf2012}. Nowadays, \emph{electrically} generated spin currents are heavily used in the field of spintronics, where they are exploited to control the direction of the magnetization in a FM layer via the spin transfer torque (STT). A similar control of the magnetization on an ultrafast timescale can be established using the \emph{optically} generated spin current, as was recently demonstrated using a non-collinear magnetic bilayer \cite{Schellekens2014,Choi2014,Choi2015}. It was argued that this optical-STT is an accurate probe of the spin current, and it will be employed in this work to investigate both the generation and absorption of the laser-pulse induced spin current.

Since its discovery, several mechanisms for the generation of the optical spin current have been suggested. Battiato et al. proposed a mechanism based on spin-dependent transport of excited electrons. In this model a superdiffusive spin current is generated due to spin filtering of the hot electrons in the FM layer \cite{Battiato2010}. A second mechanism uses the spin dependent Seebeck effect to explain the optical generation of spin currents \cite{Choi2015}. In this case the spin current is generated due to a temperature gradient across the FM material caused by the laser excitation. Lastly, there are models in which the spin current is generated by the demagnetization, for instance using the magnon-electron coupling. In this case electrons become spin polarized due to the excitation of magnons and the conservation of angular momentum, acting as a source for a diffuse spin current that follows $dM/dt$\cite{Choi2014}.

In this work, the generation of the fs laser-pulse excited spin current is investigated in order to identify which mechanism is at play. Besides the spin current generation, also the absorption depth of the spin current in a second FM layers is investigated. Both phenomena are studied using a non-collinear magnetic bilayer. This magnetic bilayer consists of one FM layer with an out-of-plane (OOP) anisotropy (generation layer), and a second FM layer with in-plane (IP) anisotropy (absorption layer). The two FM layers are separated by a metallic (non-magnetic) spacer layer. By varying the thickness of the generation layer the thickness dependence of the spin current generation is examined. It is found that the generation of spin currents is almost independent on the thickness of the generation layer. Using a wedged absorption layer the absorption of the spin current is investigated, revealing an absorption depth in Co of $2-3$ nm. It will be demonstrated that this very local absorption near the interface allows for THz spin wave excitation in these non-collinear magnetic bilayers, as was recently also demonstrated by Razdolski et al. \cite{Razdolski2016}, and being of great relevance for the upcoming field of (THz) magnonics \cite{Krawczyk2014,Chumak2015}.

\section{Sample structure and characterization}
The basic structure of the non-collinear magnetic bilayers used in this work is: SiB(substrate) / Ta(2) / Pt(4) / [Co(0.2) / Ni(0.6)]$_{N}$ / Co(0.2) / Cu(5) / Co(t$_{\mathrm{Co}}$) / Pt(1) (thickness in nm). All samples are  fabricated using DC magnetron sputtering at room temperature. In this structure the Co/Ni multilayer has an easy axis along the OOP direction (perpendicular magnetic anisotropy, PMA), and the top Co layer has an easy plane along the in-plane direction. The two FM layers are separated by a $5$ nm thick Cu layer which allows for the transfer of spin currents and decouples both FM layers. The measurements are performed using a standard time-resolved magneto-optic Kerr effect setup (TR-MOKE) in the polar configuration, and in the presence of an external field that is applied parallel to the sample surface. The probe and pump pulses have a spot size of $\approx 10 \mu$m and a pulse length of $\approx 150$ fs. The pulses are produced by a Ti:sapphire laser with a wavelength of $790$ nm and a repetition rate of 80 MHz. In the experiments the pump pulse is used to excite the spin dynamics, and the probe pulse is used to measure the time resolved OOP magnetization component of both FM layers of the non-collinear bilayer.

\begin{figure}
\includegraphics[scale=0.34,,trim = 0mm 0mm 0mm 0mm,clip]{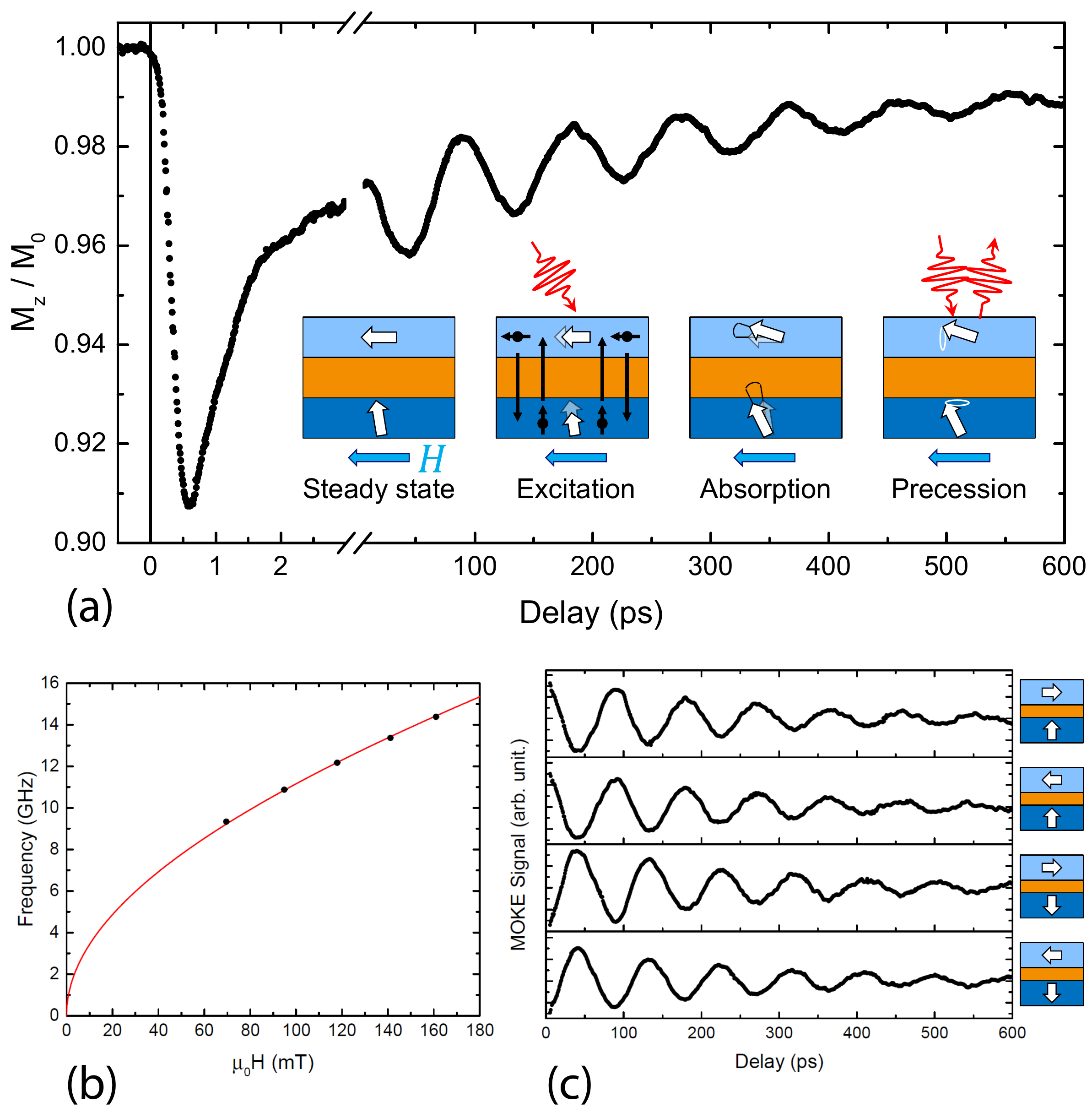}
\caption{The measurements performed on a non-collinear system with $N=4$ and $t_{\mathrm{Co}}=3$ nm. (a) Typical precession measurement with an in-plane applied field of $95$ mT. The demagnetization of the OOP layer is visible in the first picosecond, and a clear precession ($\approx 10$ GHz) is present on the long timescale. The cartoon shows the different stages before, during and after the optical excitation. (b) Precession frequency measured as a function of the in-plane applied field. The solid line represents fit using the Kittel relation, resulting in a saturation magnetization of $1.3 - 1.4$ MA.m$^{-1}$, using a surface anisotropy of $0.3 - 0.6$ mJ.m$^{-2}$ taken from literature \cite{Johnson1996} and $\gamma = 1.76 \cdot 10^{11}$ rad.s$^{-1}$.T$^{-1}$. (c) Precession measurements for all combinations of IP field direction (parallel to IP magnetization) and OOP magnetization direction (remagnetization of OOP layer subtracted from the signal).}
\label{Fig:TypicalMeasurement}
\end{figure}

The dynamics in the non-collinear system during the measurements is illustrated by the cartoon in Fig.\ \ref{Fig:TypicalMeasurement}(a). Before excitation, the system is in a \emph{steady state}, where the in-plane field sets the direction of the IP magnetization as well as a slight canting of the OOP magnetization (depending on its PMA). When the structure is \emph{excited} by the laser-pulse, spin currents are generated in both layers. These spin currents will flow through the spacer layer to be injected in the other layer. The transverse spins injected in each layer will be \emph{absorbed}, resulting in a spin-transfer-torque (STT) on the magnetization. As a result, the magnetization is canted away from the effective field, and a damped \emph{precession} is initiated. Using the probe pulse the precessions can be measured by measuring the OOP magnetization as a function of time. The amount of initial canting of the magnetization in each layer, i.e.\ the initial amplitude of the damped precession, is proportional to the absorbed angular momentum. In this work the precession amplitude of the IP layer in combination with the demagnetization of the OOP layer is used to investigate the absorption and generation of the spin current generated in the OOP layer. The precession in the OOP layer is not visible in the measurements due to the smaller MO sensitivity to the (bottom) OOP layer compared to the (top) IP layer, as will be shown in the following.

A typical measurement performed on a non-collinear bilayer with $N=4$ and $t_{\mathrm{Co}}=3$ nm is shown in Fig.\ \ref{Fig:TypicalMeasurement}(a). In this figure the OOP magnetization of both layers, normalized to the magnetization of the OOP layer, is plotted as a function of the pump-probe delay. In the first few picoseconds the demagnetization and subsequent remagnetization of the OOP layer is visible. On the long timescale a clear precession of $\approx 10$ GHz is observed. As is illustrated in the cartoon of Fig.\ \ref{Fig:TypicalMeasurement}(a), the optical generated spin currents are expected to induce a precessional motion of both the IP and OOP magnetization. Moreover, a precession in both FM layers was measured in previous experiments by Schellekens et al. \cite{Schellekens2014}, using similar structures. In their work it was demonstrated that the precession of the IP magnetization was indeed initiated by the \emph{STT mechanism} discussed earlier. The precession of the OOP magnetization, however, was attributed to a laser-pulse induced anisotropy change, which was called the \emph{$\Delta K$ mechanism} and will be discussed in more detail later. In this work the precession in the IP layer will be used to measure the absorption and generation of the spin current generated in the OOP layer. Therefore, it needs to be confirmed that the measured precession actually is the precessional motion of the IP magnetization, initiated by the STT mechanism. In the following it will be shown that this is indeed the case.

One way to identify to which layer the precession corresponds is to perform field dependent precession measurements. In case of the IP magnetization in an in-plane external field the frequency $f_{\mathrm{IP}}$ is given by the Kittel relation
\begin{equation}
    f_{\mathrm{IP}} = \frac{\gamma}{2 \pi} \sqrt{B_{\mathrm{app}} \left( B_{\mathrm{app}} + \mu_{0} M_{\mathrm{s}} - \frac{2 K_{\mathrm{s}}}{t M_{\mathrm{s}}} \right)},
\label{Eq:KittelIP}
\end{equation}
in which $\gamma$ is the gyromagnetic ratio, $B_{\mathrm{app}}$ is the applied field, $t$ and $M_{\mathrm{s}}$ the thickness and saturation magnetization of the magnetic layer and $K_{\mathrm{s}}$ the surface anisotropy constant (including the contribution of both interfaces of the FM layer). In case of the OOP magnetization in an in-plane field, where the applied field is small compared to the anisotropy field, the precession frequency $f_{\mathrm{OOP}}$ is given by
\begin{equation}
    f_{\mathrm{OOP}} = \frac{\gamma}{2 \pi} \sqrt{\left( \mu_{0} M_{\mathrm{s}} - \frac{2 K_{\mathrm{s}}}{tM_{\mathrm{s}}} \right)^{2} - B_{\mathrm{app}}^{2}}.
\label{Eq:KittelOOP}
\end{equation}
From these equations it is seen that in case of the IP (OOP) magnetization the frequency increases (decreases) when the applied field is increased. Figure\ \ref{Fig:TypicalMeasurement}(b) shows the measured frequency as a function of the applied field (black dots). A clear increase of the frequency with field is observed. Moreover, the field dependence is nicely fitted with the Kittel relation of Eq.\ (\ref{Eq:KittelIP}) (red curve), resulting in a saturation magnetization of $1.3 - 1.4$ MA.m$^{-1}$, using a surface anisotropy of $0.3 - 0.6$ mJ.m$^{-2}$ derived from literature \cite{Johnson1996} and $\gamma = 1.76 \cdot 10^{11}$ rad.s$^{-1}$.T$^{-1}$. The found saturation magnetization compares well with the bulk value for Co of $1.4$ MA.m$^{-1}$, unambiguously demonstrating that it is the precessional motion of the IP magnetization that is measured.

Next, it should be verified that the mechanism initiating the precession is indeed the STT mechanism, and not the earlier mentioned $\Delta K$ mechanism. The latter mechanism arises when the field is applied at a certain angle to the sample surface, i.e.\ due to a minor misalignment \cite{Kampen2002}. With the field at an angle, the equilibrium direction of the effective field and thus the magnetization is no longer IP, but is canted slightly out of plane. A precession can be initiated by a laser-pulse induced change in the magnetization and anisotropy, abruptly altering the effective field direction and resulting in a precession of the magnetization. Fortunately, a distinction between the two mechanisms can be made by looking at the sign of the measured precession when the applied field and OOP magnetization directions are inverted. In case of the $\Delta K$ mechanism, the precession signal inverts with the field direction, but is independent of the OOP magnetization direction. On the contrary, for the STT mechanism, the precession signal is independent of the field direction, and is inverted when the the magnetization direction of the OOP layer reverses. Figure\ \ref{Fig:TypicalMeasurement}(c) shows the precession measured for all combinations of IP field direction (parallel to IP magnetization) and OOP magnetization direction. Looking at the top two curves it can be seen that the precession is identical for both field directions. The precession signal inverts when the OOP magnetization direction is reversed, as can be seen in the bottom two curves. This confirms that the measured precession of the IP magnetization is indeed initiated by the STT mechanism.

It was noted before that in previous experiments on similar structures performed by Schellekens et al., also a precession of the OOP magnetization was measured, which was initiated by the $\Delta K$ mechanism \cite{Schellekens2014}. The absence of this precession in the present measurements is caused by the addition of a Ta seed layer underneath the Co/Ni multilayer, causing a strong increase of the PMA and the corresponding anisotropy field. With an increase of the anisotropy field of the OOP layer the effect of the laser-pulse excitation on the effective field becomes smaller, reducing the amplitude of the $\Delta K$ precession to a point where it is not measurable anymore. The exclusion of the precession of the OOP magnetization from the measurements allows for a more straightforward analysis of the measured data.

With the measured spin dynamics verified, it can be used to investigate the generation and absorption of the laser-pulse excited spin current. To do so, two parameters are defined, being the \emph{efficiency}, $\epsilon$, and the \emph{initial canting angle}, $\theta_{\mathrm{c}}$. The efficiency is defined as the ratio of OOP angular momentum absorbed by the IP layer, $\Delta M_{z,\mathrm{IP}}$, to the angular momentum lost during demagnetization by the OOP layer, $\Delta M_{z,\mathrm{OOP}}$,
\begin{equation}
    \epsilon = \frac{\Delta M_{z,\mathrm{IP}}}{\Delta M_{z,\mathrm{OOP}}}.
\label{Eq:Efficiency}
\end{equation}
The initial canting angle is defined as the angle of the IP magnetization with respect to the sample surface after absorption of the OOP spin current, and can be calculated using
\begin{equation}
    \theta_{\mathrm{c}} = \arcsin \left( \frac{\Delta M_{z,\mathrm{IP}}}{M_{\mathrm{s,IP}} t_{\mathrm{IP}}} \right).
\label{Eq:CantingAngle}
\end{equation}
In this equation $M_{\mathrm{s,IP}}$ and $t_{\mathrm{IP}}$ are the saturation magnetization and thickness of the IP layer, respectively. A discussion on how the efficiency and canting angle are derived from a precession measurement as shown in Fig.\ \ref{Fig:TypicalMeasurement}(a) can be found in Appendix\ \ref{app:CalculationEff}.

\begin{figure}
\includegraphics[scale=0.3,trim = 0mm 0mm 0mm 0mm,clip]{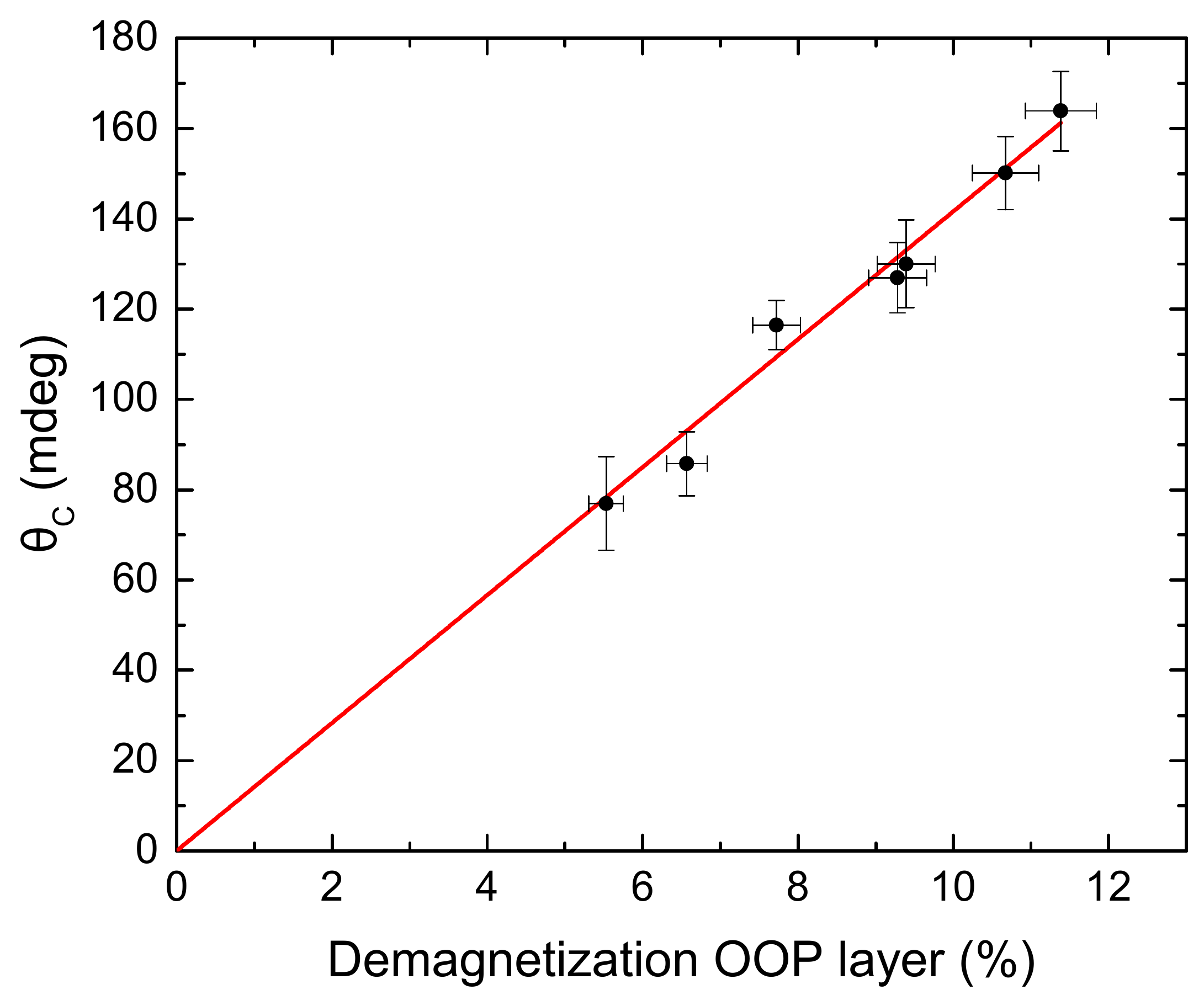}
\caption{Fluence dependent measurement performed on a non-collinear system with $N=4$ and $t_{\mathrm{Co}}=3$ nm. The figure shows the measured canting angle of the IP magnetization as a function of the demagnetization of the OOP layer. The solid line represents a linear fit with fixed zero offset.}
\label{Fig:PowerDependence}
\end{figure}

It is important to note that measurements on different samples or different areas on the same sample are going to be compared. The amount of demagnetization in these different measurements might be slightly different, e.g.\ due to a small difference in spotsize or pump-probe overlap. In order to be able to compare the different measurements, care should be taken that the measured parameters are independent of the demagnetization of the OOP layer. This is investigated by measuring $\theta_{\mathrm{c}}$ as a function of the demagnetization of the OOP layer for a non-collinear system with $N=4$ and $t_{\mathrm{Co}}=3$ nm. The results of this measurement are presented by the black dots in Fig.\ \ref{Fig:PowerDependence}. The figure shows that for the low demagnetization regime used throughout this work $\theta_{\mathrm{c}}$ is linear in the amount of demagnetization (red line). The achieved $\theta_{\mathrm{c}}$ is on the order of millidegrees, and thus can be approximated by $\Delta M_{z,\mathrm{IP}} / \left( M_{\mathrm{s,IP}} t_{\mathrm{IP}} \right)$. This means that $\Delta M_{z,\mathrm{IP}}$ scales linearly with the demagnetization, i.e.\ with $\Delta M_{z,\mathrm{OOP}}$, and thus that the efficiency is independent of the amount of demagnetization. Moreover, the linear dependence of $\theta_{\mathrm{c}}$ shown in Fig.\ \ref{Fig:PowerDependence} also means that the canting angle per percent demagnetization, $\theta_{\mathrm{c,\%}}$, can be used as a demagnetization independent parameter. In conclusion, both $\epsilon$ and $\theta_{\mathrm{c,\%}}$ are good parameters to be compared between different measurements.

\section{Results and discussion}
\subsection{Spin current absorption and THz spin wave excitation}
First the absorption of the spin current in the IP layer is investigated. By using a wedge shaped top Co layer the penetration depth of the transverse spins is measured. The structure used in this measurement is given by the basic non-collinear bilayer introduced earlier, now with $N=4$ and a wedge shaped top Co layer with $t_{\mathrm{Co}}$ ranging from $0$ nm to $6$ nm over a distance of $20$ mm. Using the fact that the TR-MOKE measurement is a very local technique (spotsize $\approx 10$ $\mu$m), the thickness dependent measurement can be performed by measuring $\epsilon$ and $\theta_{\mathrm{c,\%}}$ at different points along the Co wedge.

\begin{figure}
\includegraphics[scale=0.3,trim = 0mm 0mm 0mm 0mm,clip]{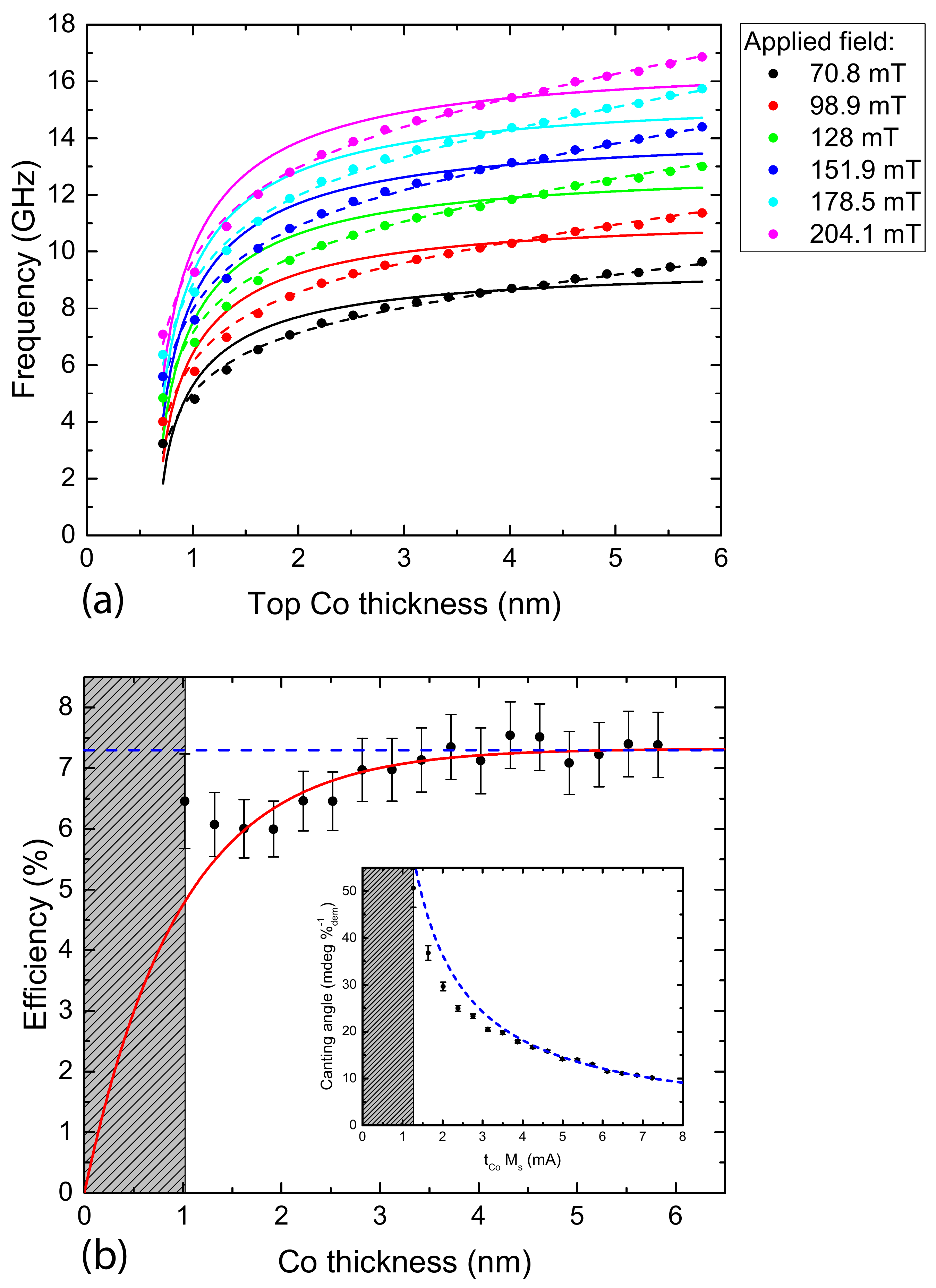}
\caption{Measurements performed on a non-collinear magnetic bilayer with $N = 4$ and wedged top Co layer with thickness ranging from $0$ nm to $6$ nm over a distance of 20 mm. (a) Precession frequency measured as a function of top Co layer thickness for six different in-plane applied field strengths. The solid curves represent fits using the standard Kittel relation, resulting in $M_{\mathrm{s,IP}} \approx 1240$ kA.m$^{-1}$ and $K_{\mathrm{s}} \approx 0.70$ mJ.m$^{-2}$. The dotted curves represent fits using the Kittel relation including a thickness dependent saturation magnetization as discussed in Appendix\ \ref{app:CalculationLinMs} (b) Efficiency and canting angle per percent demagnetization (inset) as a function of top Co layer thickness. The solid curve represents a fit to the data showing a finite absorption depth (e.g.\ $90$ \% absorbed within first $2.2 \pm 0.2$ nm). The dotted lines represent constant efficiency and corresponding $\theta_{\mathrm{c,\%}}$, which describes the case when there is full absorption independent of top Co layer thickness.}
\label{Fig:SCAbsorption}
\end{figure}

The saturation magnetization and surface anisotropy of the wedged IP layer, needed for the determination of $\epsilon$ and $\theta_{\mathrm{c,\%}}$, are obtained by measuring the frequency of the precession as a function of the IP layer thickness. The Kittel relation (Eq.\ (\ref{Eq:KittelIP})) shows that indeed the thickness dependence of the anisotropy term allows to determine both $M_{\mathrm{s,IP}}$ and $K_{\mathrm{s}}$ from the measured data. The measured frequencies as a function of the Co layer thickness, for different applied field strengths, are shown by the solid dots in Fig.\ \ref{Fig:SCAbsorption}(a). The solid curves are fits to the data using Eq.\ (\ref{Eq:KittelIP}). The fits are performed using a global fit with shared fitting parameters $M_{\mathrm{s,IP}}$ and $K_{\mathrm{s}}$, resulting in $M_{\mathrm{s,IP}} \approx 1240$ kA.m$^{-1}$ and $K_{\mathrm{s}} \approx 0.70$ mJ.m$^{-2}$. Looking at the fitted curves, it can be seen that the measured data are not well described by the Kittel relation. For all field strengths it seems that there is an additional thickness dependence that is not captured by Eq.\ (\ref{Eq:KittelIP}). As will be discussed later, one possibility is a thickness dependent saturation magnetization, decreasing for thinner layer thicknesses. For simplicity, however, the simplest case with constant $M_{\mathrm{s,IP}}$ and $K_{\mathrm{s}}$ throughout the wedged layer will be used in the remainder of this discussion. The analysis including a thickness dependent $M_{\mathrm{s,IP}}$ can be found in Appendix\ \ref{app:CalculationLinMs} (results are shown by the dashed curves in Fig.\ \ref{Fig:SCAbsorption}(a)). There it is shown that the overall behaviour of the efficiency and canting angle as a function of the Co layer thickness is robust and similar for both cases.

The efficiency calculated as a function of the top Co layer thickness is shown in Fig.\ \ref{Fig:SCAbsorption}(b) (solid dots). For the very thin Co thicknesses (grey area) no precession amplitudes could be determined, which is attributed to the surface anisotropy (PMA) becoming too pronounced. The solid curve is a fit to the data following
\begin{equation}
\epsilon = \epsilon_{\mathrm{max}} \left( 1 - e^{-\frac{t_{\mathrm{Co}}}{\lambda_{t,\mathrm{Co}}}} \right),
\label{Eqn:ExpFitFunction}
\end{equation}
where $\epsilon_{\mathrm{max}}$ is the efficiency for infinite Co thickness $t_{\mathrm{Co}}$, and $\lambda_{\mathrm{Co}}$ the characteristic spin absorbtion length. For large thicknesses the efficiency saturates, corresponding to full absorption of the transverse spin current. At zero thickness, i.e.\ no IP layer, the efficiency must be zero. Looking back at the definition of the efficiency (Eq.\ (\ref{Eq:Efficiency})), it can be seen that the behaviour shown in Fig.\ \ref{Fig:SCAbsorption}(b) corresponds to a transverse spin absorption that decays exponentially with the distance from the interface where the spins are injected. From the fit the values $\lambda_{\mathrm{Co}} = 0.96 \pm 0.07$ nm and $\epsilon_{\mathrm{max}} = 7.3 \pm 0.1$ \% are obtained. This result shows that the spin current is absorbed very locally near the interface, i.e.\ $90$ \% of the transverse spins are absorbed within the first $2.2 \pm 0.2$ nm of the Co layer. This penetration depth agrees well with the penetration depth of $1.7$ nm found for electrically driven transverse spin currents in Co by Ref.\ \cite{Taniguchi2008}.

The effect of the limited penetration depth of the transverse spins is also seen in the canting angle per percent demagnetization, as shown in the inset of Fig.\ \ref{Fig:SCAbsorption}(b). Here $\theta_{\mathrm{c,\%}}$ is plotted as a function of $t_{\mathrm{Co}} M_{\mathrm{s}}$ (solid dots). For small canting angles a $(t_{\mathrm{Co}} M_{\mathrm{s}})^{-1}$ dependency is expected when there is full absorption, i.e.\ constant efficiency ($\lambda_{\mathrm{Co}} \rightarrow 0$ nm), illustrated by the dotted curve. It is seen that for Co thicknesses below approx. $3$ nm the canting angle is not reaching its maximum value, demonstrating again the incomplete absorption of the spin current for these thicknesses.

\begin{figure}
\includegraphics[scale=0.3,trim = 0mm 0mm 0mm 0mm,clip]{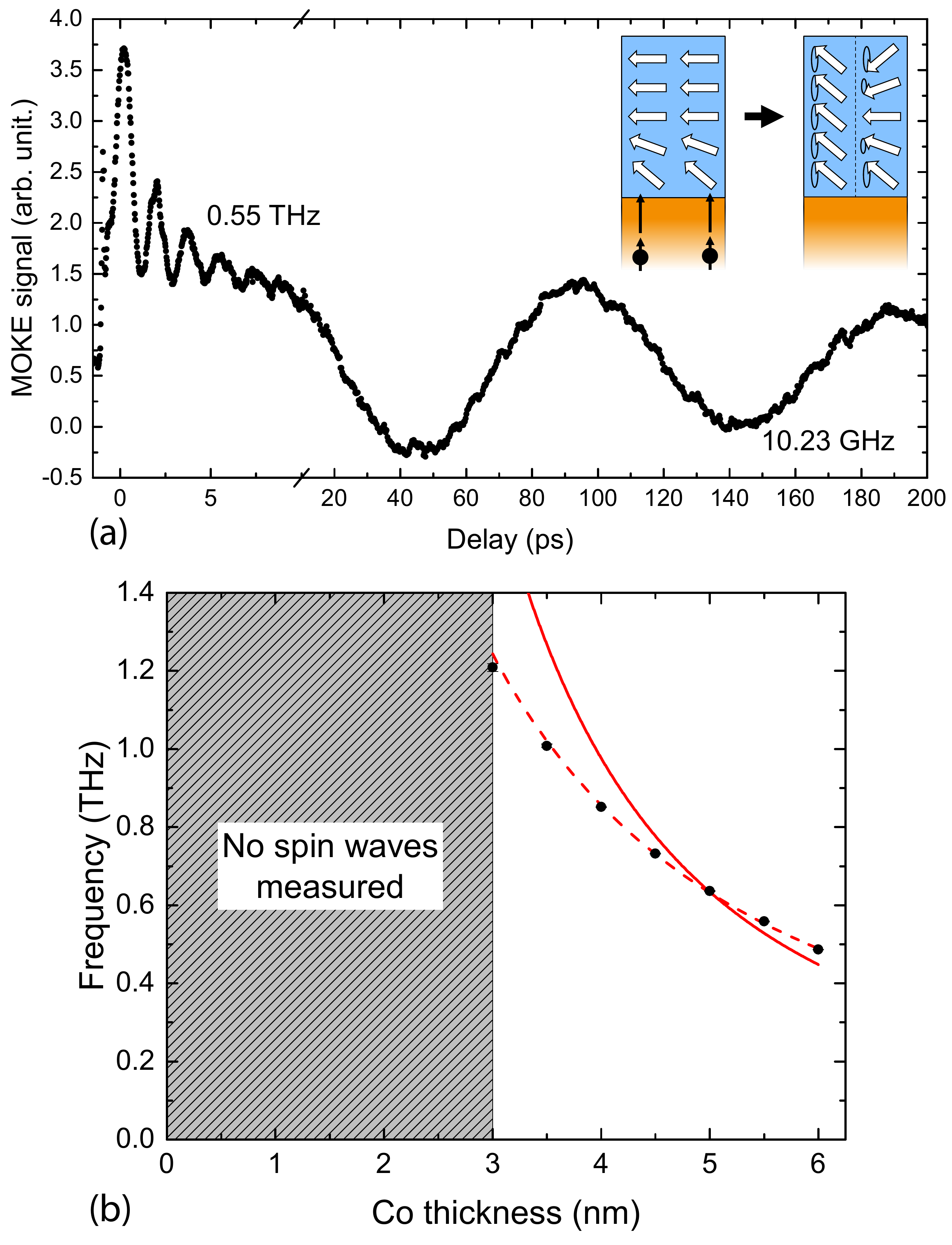}
\caption{(a) Precession measurement performed on a non-collinear bilayer with $N = 4$ and wedged top Co layer with a thickness ranging from $0$ nm to $6$ nm, measured at a thickness of $5.5$ nm. During the measurement a QWP was added to TR-MOKE setup to increase the sensitivity to the THz precession. Both the ($\approx 0.55$ THz) spin wave and the ($\approx 10$ GHz) fundamental precession are visible. Cartoon shows the gradient in the canting angle within the Co layer after the optical excitation, and the resulting fundamental and first order standing spin wave. (b) Spin wave frequency measured as a function of the Co layer thickness. No THz precessions are found for Co thicknesses below $3$ nm. The dotted curve represents a fit using the standard dispersion relation, showing a clear discrepancy with the measured data. The solid curve represents a fit using the dispersion relation with a thickness dependent spin wave stiffness, showing a much more accurate description of the data.}
\label{Fig:THzSpinWaves}
\end{figure}

The results on the spin absorption shows that the absorption of the transverse spins falls of exponentially with the distance from the injection interface. This results in a strong gradient in the canting angle of the IP magnetization, as illustrated in the left cartoon of Fig.\ \ref{Fig:THzSpinWaves}(a). It was recently demonstrated by Razdolski et al. that a strong gradient in the magnetization direction can be used to excite THz standing spin waves along the depth of the IP layer \cite{Razdolski2016}. In the following it will be demonstrated that these THz spin waves, with frequencies up to $1.2$ THz, are indeed excited in the non-collinear bilayer measured here.

The THz spin waves are observed as an additional precession on the fast picosecond timescale in the demagnetization measurements performed in the thicker region of the Co wedge (not shown). The spin waves carry no net OOP magnetic moment, causing their signal to be averaged out in case of homogeneous averaging across the thickness of the layer. The fact that the spin waves can be measured in the TR-MOKE setup results from a certain depth sensitivity due to the attenuation of the laser within the FM layer. However, for these thin layers the depth sensitivity and thereby the spin wave signal is only small. To achieve a better sensitivity of the TR-MOKE setup to the THz precession a quarter-wave-plate (QWP) was added to the probe beam. By carefully tuning the QWP angle, a specific linear combination of Kerr rotation and ellipticity can be measured \cite{Schellekens2014_2}, and the signal can be optimized to measure the THz precession. An example of such a measurement is shown in Fig.\ \ref{Fig:THzSpinWaves}(a). In this figure both the first order standing spin wave (0.55 THz) as well as the fundamental precession (10 GHz) are visible (different timescales), both illustrated in the right cartoon in the figure. The shown measurement is performed with an IP layer thickness of $t_{\mathrm{Co}} = 5.5$ nm. The sign of the THz precession has the same dependency on the IP and OOP magnetization direction as the fundamental precession as was shown in Fig.\ \ref{Fig:TypicalMeasurement}(c). Also, the THz precession is even present without the applied field, which is expected with the exchange interaction driving the precession.

Using the wedge shape of the IP layer in the non-collinear bilayer the spin wave frequency can be measured as a function of the Co layer thickness. The frequencies measured for the different Co thicknesses are shown by the black dots in Fig.\ \ref{Fig:THzSpinWaves}(b). At thicknesses of $2.5$ nm and below there was no sign of the spin wave in the measurement. This means that either the spin waves are not excited, or that they are not visible in the measurement. At the moment it is believed that the latter is the case, being the result of a strong decrease in the lifetime of the spin waves due to the large increase in frequency, combined with a decrease in the depth sensitivity of the MOKE for the thinner layers.

The thickness dependence of the spin wave frequency shown in Fig.\ \ref{Fig:THzSpinWaves}(b) can be fitted using the theoretical dispersion relation in order to obtain the spin wave stiffness of the Co layer. The dispersion relation, including both the in-plane applied field ($B_{app}$) and shape and surface anisotropy terms contributions, is given by (see Appendix\ \ref{app:CalculationDispersion} for derivation)
\begin{multline}
f\left(k\right) = \frac{\gamma}{2 \pi} \left[\left(B_{\mathrm{app}} + \frac{D_{\mathrm{sw}}}{\gamma \hbar} k^{2}\right) \right. \\ \left.\left(B_{\mathrm{app}} + \mu_{0} M_{\mathrm{s}} - \frac{2 K_{\mathrm{s}}}{t M_{\mathrm{s}}} + \frac{D_{\mathrm{sw}}}{\gamma \hbar} k^{2}\right)\right]^{\frac{1}{2}},
\label{Eq:Dispersion}
\end{multline}
\begin{equation}
k = \frac{\pi n}{t}.
\end{equation}
In this equation the spin wave frequency and order are given by $f$ and $n$, respectively. The spin wave stiffness is represented by $D_{\mathrm{sw}}$, and $\hbar$ corresponds to the reduced Planck constant.  Using this relation, with $n = 1$, $M_{\mathrm{s}} = 1240$ kA.m$^{-1}$, $K_{\mathrm{s}} = 0.70$ mJ.m$^{-2}$ and $B_{\mathrm{app}} = 72$ mT, the data in Fig.\ \ref{Fig:THzSpinWaves}(b) can be fitted, using the spin wave stiffness as the fitting parameter. Looking at the fitted curve (solid red curve), it can be seen that the measured data is not well described by the dispersion relation of Eq.\ (\ref{Eq:Dispersion}). The measured dispersion is flattened out with respect to the theoretical dispersion. This suggests, as was seen for the Kittel fits in Fig.\ \ref{Fig:SCAbsorption}(a), that here is an additional thickness dependence that is not captured by the dispersion relation in Eq.\ (\ref{Eq:Dispersion}). A detailed investigation of this additional thickness dependence is out of the scope of this paper, and only a short discussion on a possible explanation will be given.

In the case of $n > 0$, the dispersion relation shown in Eq.\ (\ref{Eq:Dispersion}) is dominated by the  $D_{\mathrm{sw}}$ term. Therefore, the additional thickness dependence can be expected to be present in this spin wave stiffness. The spin wave stiffness itself is related to the exchange constant $A_{\mathrm{ex}}$ via the atomic spin $S$ and the lattice constant $a$,
\begin{equation}
D_{\mathrm{sw}} = \frac{A_{\mathrm{ex}} a^{3}}{S}.
\label{Eq:DswAex}
\end{equation}
It has been demonstrated by Enrich et al. that the exchange constant in Co decreases when it is alloyed with other materials \cite{Eyrich2014}. Moreover, they demonstrated a significant decrease of the average exchange constant of the Co layers in a Co/Ru multilayer for Co thicknesses below $10$ nm. The decrease of $A_{\mathrm{ex}}$ was attributed to intermixed interface regions that have a lower exchange constant, which become more dominant for the thinner Co layers. In case of the sputter deposited structures measured here, interface intermixing is expected to occur as well. Since the data presented in Ref.\ \cite{Eyrich2014} suggest an exponential dependence of $A_{\mathrm{ex}}$ on the Co thickness, the following estimation is used,
\begin{equation}
D_{\mathrm{sw}}(t) = D_{\mathrm{sw},\infty}\left(1-e^{-\frac{t}{d_{0}}}\right).
\label{Eq:DswExp}
\end{equation}
Substituting Eq.\ (\ref{Eq:DswExp}) into the dispersion relation, the measured spin wave frequencies are fitted with much more accuracy, as is shown by the dotted curve in Fig.\ \ref{Fig:THzSpinWaves}(b). The fitted parameters are $D_{\mathrm{sw},\infty} = 980 \pm 30$ meV.\AA$^{2}$ and $d_{0} = 4.7 \pm 0.3$ nm. The $d_{0}$ value seems to be reasonable compared to the data on $A_{\mathrm{ex}}$ in Ref.\ \cite{Eyrich2014} (estimated value of $d_{0} \approx 4.3$ nm). However, the value for $D_{\mathrm{sw},\infty}$, i.e.\ for bulk Co, is almost a factor of three higher than the expected value of $340 \pm 75$ meV.\AA$^{2}$ found for poly-crystalline Co \cite{Vernon1984}. This discrepancy suggests that the description of the spin wave stiffness as given in Eq.\ (\ref{Eq:DswExp}) is not complete and a more elaborate study is needed, which is out of the scope of this paper.

The thickness dependent exchange constant being the cause of the observed (flattened) THz dispersion is a first assumption and needs to be confirmed, however, it does also give a possible explanation for the discrepancies observed for the fundamental precessions shown in Fig.\ \ref{Fig:SCAbsorption}(a). The Curie temperature of a magnetic layer is directly proportional to the exchange constant, meaning that a lowering in the exchange constant results in a decrease of the Curie temperature. Since the measurements are performed at room temperature, this decreases the saturation magnetization and with it the precession frequency. As was mentioned earlier, adding such a thickness dependent $M_{\mathrm{s}}$ to Eq.\ (\ref{Eq:KittelIP}) indeed results in a better fit to the measured data (analysis in Appendix\ \ref{app:CalculationLinMs}). This shows that the additional thickness dependence observed in the dispersion of both the fundamental precessions as well as the spin waves can be explained with a thickness dependence in the average exchange constant of the Co film, possibly caused by intermixing in the interface regions.

\subsection{Spin current generation}
Next, the laser-pulse excited spin current generation in the OOP layer is investigated. This is done by measuring the efficiency and canting angle as a function of the OOP layer thickness. For this measurement four structures are fabricated following the basic non-collinear bilayer as introduced earlier. Each structure has a similar top Co layer with $t_{\mathrm{Co}} = 3$ nm, but a different amount of [Co/Ni]$_{N}$ repeats. The used repeats are $N = 1, 2, 3$ and $4$. Unfortunately, the measurements on the structure with $N = 1$ showed no precessions of the IP layer. Polar hysteresis measurements on this structure revealed a very small coercivity for the OOP layer ($\approx 1$ mT), indicating weak PMA. As a result the OOP layer is pulled in-plane by the applied field during the measurements. In this case parallel spins are injected in the IP layer and there will be no STT and thus no canting of the magnetization. This is seen in the measurements by the absence of both the demagnetization and the precession. Although this causes the structure to be useless for the spin current investigation, it confirms again that the measured precession of the IP magnetization is indeed caused by the STT mechanism.

\begin{figure}
\includegraphics[scale=0.3,trim = 0mm 0mm 0mm 0mm,clip]{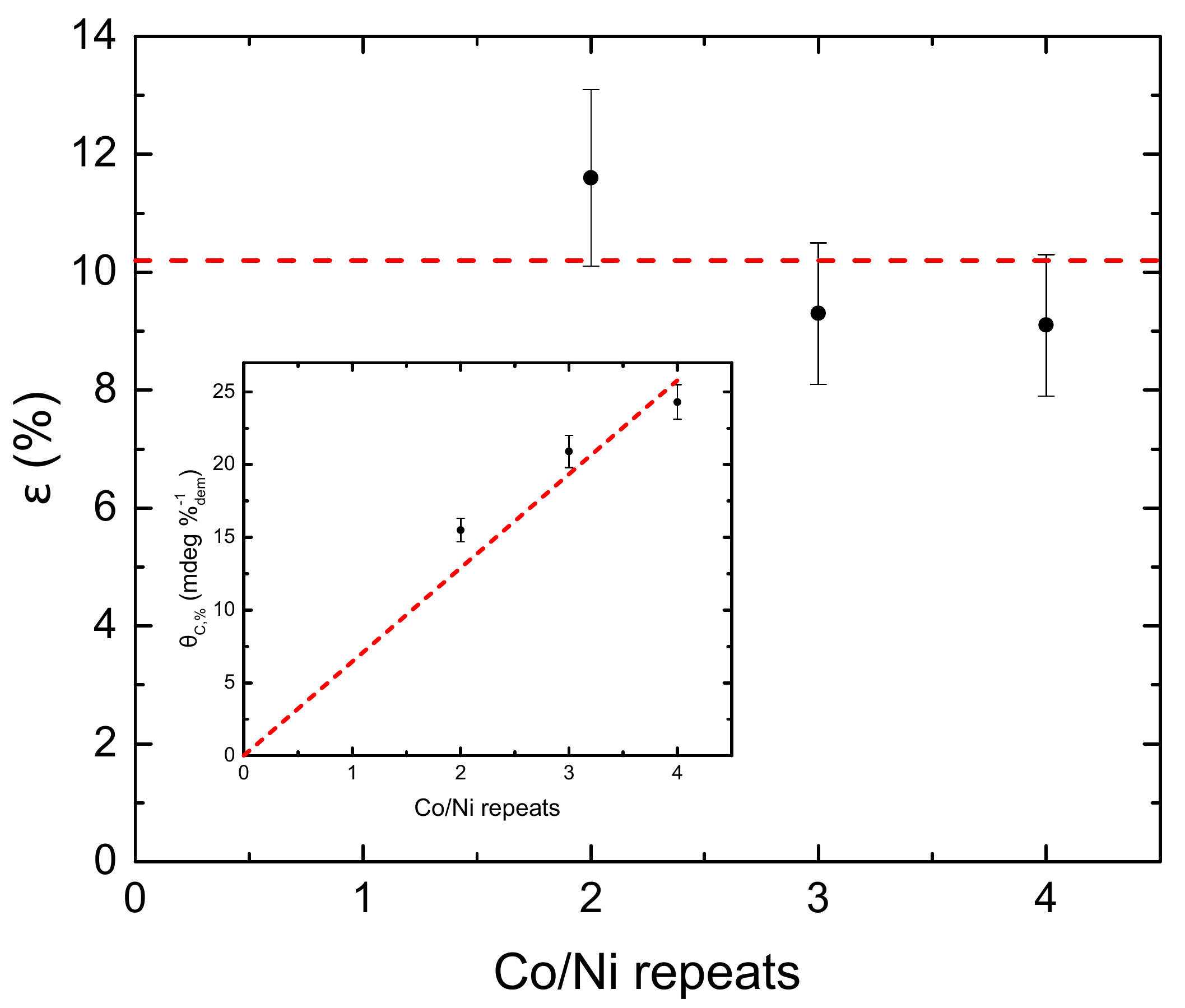}
\caption{Efficiency and canting angle per percent demagnetization (inset) measured on the non-collinear bilayers with $t_{\mathrm{Co}} = 3$ nm and [Co/Ni]$_{N}$ repeats of $N = 2,3$ and $4$. For the efficiency no significant dependence on the amount of repeats is seen, whereas for the canting angle per percent demagnetization a clear increase with number of repeats is present. The dotted lines are guides to the eye.}
\label{Fig:SCGeneration}
\end{figure}

The samples with $N = 2, 3$ and $4$ did show sufficient PMA and the efficiency and canting angle could be determined. The resulting $\epsilon$ and $\theta_{\mathrm{c,\%}}$ as function of the [Co/Ni]$_{N}$ repeats are shown in Fig.\ \ref{Fig:SCGeneration} (dotted lines are guides to the eye). First looking at the efficiency, it can be seen that there seems to be no strong dependency on the thickness of the OOP layer. A constant efficiency would mean that for an equal amount of angular momentum lost in the OOP layer, an equal amount of angular momentum is absorbed by the IP layer, independent of the thickness of the OOP layer. This behaviour can also be seen in the measured $\theta_{\mathrm{c,\%}}$ as a function of [Co/Ni]$_{N}$ repeats, shown in the inset. The amount of angular moment lost in the OOP layer per percent demagnetization increases with the amount of repeats. With a constant $\epsilon$, the amount of angular momentum absorbed in the IP layer will also increase with the [Co/Ni]$_{N}$ repeats, resulting in the rise of $\theta_{\mathrm{c,\%}}$. Although, due to the limited number of thicknesses available here, a more extensive thickness study is needed to confirm the observed behaviour of both $\epsilon$ and $\theta_{\mathrm{c,\%}}$, this observation does allow for a first speculation on the spin current generation mechanism.

It is noted that the stacks (spacer layer and top Co layer) deposited on top of the different [Co/Ni]$_{N}$ repeats are designed to be identical. This means that the transport and absorption of the spin current generated in the OOP layer is expected to be the same in all three structures. The absence of a clear thickness dependence in $\epsilon$ therefore indicates that a similar amount of angular momentum loss in the OOP layer generates a similar spin current, independent of the thickness of the OOP layer. Also, it implies that the full thickness of the OOP layer contributes to the generated spin current. The latter notion contradicts the idea of a limited interface region ($\approx 1$ nm) contributing to the spin current generation, as was suggested by Alekhin et al. for Fe/Au \cite{Alekhin2014}. In case of a superdiffusive spin current the spin current is generated due to spin filtering of the hot electrons in the magnetic layer \cite{Battiato2010}. Although for a complete assessment explicit calculations need to be performed, it might be expected that the spin filtering becomes more pronounced with increasing layer thickness since part of the hot electrons will have to travel a longer distance within the FM layer. In that case the net spin current leaving the OOP layer would increase with the layer thickness, resulting in an increase of the efficiency with [Co/Ni]$_{N}$ repeats. This, however, is not observed in the present measurement. The spin current generated by the spin dependent Seebeck effect is negligible in the structures used in this work, as was demonstrated by Schellekens et al. for similar non-collinear bilayers \cite{Schellekens2014}. The notion of the generated spin current being solely dependent on the amount of lost angular momentum does agree with a mechanism where the spin current is generated by the demagnetization, thus following $dM/dt$\cite{Choi2014}. In this case a certain amount of loss in angular momentum in the OOP layer ($dM$) will generate a certain (diffuse) spin current, independent of the thickness of the OOP layer. While this analysis tentatively points towards a $dM/dt$ like scenario, the main conclusion is that the type of experiments presented can be highly valuable to resolve the optical-STT mechanism. Explicit model calculations for the different scenarios need to be performed before making an unambiguous assignment.

\section{Conclusion}

In conclusion, both the generation and absorption of fs laser-pulse induced spin currents have been experimentally investigated using non-collinear magnetic bilayers. Using a wedge shaped Co (absorption) layer it has been demonstrated that the spin current is absorbed very locally near the injection interface ($90\%$ within the first $\approx 2$ nm). This local absorption was confirmed by the demonstration of THz spin waves being excited within the Co layer as a result of the strong gradient in the canting angle of the IP magnetization after the optical excitation. Also, the mechanism behind the optical spin current generation in these magnetic bilayers has been examined. This was done by measuring the spin current generation as a function of the Co/Ni (generation) layer thickness. The results indicate that the spin current generation is solely dependent on the amount of lost angular momentum, and not on the thickness of the layer, favouring a mechanism where the spin current is generated by the demagnetization, and follows $dM/dt$. The experiments presented in this work demonstrate that the non-collinear bilayer is a convenient structure to investigate optically generated spin current. Moreover, the possibility to excite THz spin waves in these structures causes them to be of high potential for future THz magnonics.

\acknowledgments{This work is part of the Gravitation program 'Research Centre for Integrated Nanophotonics', which is financed by the Netherlands Organisation for Scientific Research (NWO).}

\bibliographystyle{unsrt}
\bibliography{Literaturelist}

\appendix
\section{}
\label{app:CalculationEff}

In this section it will be discussed how the efficiency and canting angle are derived from the precession measurements, of which a typical example is shown in Fig.\ \ref{Fig:TypicalMeasurement}(a). Since an elaboration on the calculation can already be found in Ref. \cite{Schellekens2014}, only a brief overview will be given here.

The efficiency is defined as the ratio of OOP angular momentum absorbed by the IP layer to the angular momentum lost in the OOP layer during demagnetization. This ratio is determined using the ratio of the precession amplitude of the IP layer, $A_{\mathrm{osc}}$, to the the amplitude of the demagnetization of the OOP layer, $A_{\mathrm{dem}}$. Both amplitudes are obtained from the precession measurement as shown in Fig.\ \ref{Fig:TypicalMeasurement}(a). This ratio needs to be multiplied with a sensitivity factor, $f_{\mathrm{MO}}$, to correct for the difference in magneto-optical (MO) sensitivity to the spins in either magnetic layer. Thus, the efficiency is given by
\begin{equation}
    \epsilon = \frac{\Delta M_{z,\mathrm{IP}}}{\Delta M_{z,\mathrm{OOP}}} = \frac{A_{\mathrm{osc}}}{A_{\mathrm{dem}}} f_{\mathrm{MO}}.
\label{Eq:Efficiency2}
\end{equation}
In order to determine the sensitivity factor (static) hysteresis measurements are performed using the polar MOKE setup and an out-of-plane applied external field. This measurement shows an easy axis switch for the OOP magnetization on top of a linear background signal corresponding to the hard axis rotation of the IP magnetization. The MO signal of the OOP layer is simply given by the stepsize of the switch, $A_{\mathrm{OOP}}$. In case of the IP layer the MO signal is calculated using the slope of the hard axis rotation, $B_{\mathrm{IP}}$, and the calculated saturation field, $\mu_{0}H_{\mathrm{sat}}$. The sensitivity factor is than calculated by taking the ratio of the MO signals of each layer normalized to its magnetic moment $M$,
\begin{equation}
    f_{\mathrm{MO}} = \frac{A_{\mathrm{OOP}}/M_{\mathrm{OOP}}}{B_{\mathrm{IP}} \mu_{0}H_{\mathrm{sat}}/M_{\mathrm{IP}}}.
\label{Eq:fMO}
\end{equation}
The magnetic moment of the OOP layer is determined using a SQUID-VSM, whereas the magnetic moment of the IP layer is determined from the field dependent precession frequency.

The OOP magnetic moment absorbed by the IP layer is given by the precession amplitude and the sensitivity factor, multiplied by the magnetic moment of the OOP layer due to the normalization of the precession measurement: $\Delta M_{z,\mathrm{IP}} = A_{\mathrm{osc}}f_{\mathrm{MO}}M_{\mathrm{OOP}}$. This means that the canting angle can be calculated using
\begin{equation}
    \theta_{\mathrm{c}} = \arcsin \left( \frac{A_{\mathrm{osc}}M_{\mathrm{OOP}}f_{\mathrm{MO}}}{M_{\mathrm{IP}}} \right).
\label{Eq:CantingAngle2}
\end{equation}

It is noted that in the present work care is taken that the calculated slope of the hard axis rotation of the IP magnetization is corrected for the linear background signal resulting from the Faraday effect in the optical components of the polar MOKE setup, and that the surface anisotropy is taken into account in the calculation of the saturation field of the IP magnetization.

\section{}
\label{app:CalculationDispersion}

In this section the dispersion relation of the standing spin waves (Eq.\ (\ref{Eq:Dispersion})) will be derived. In this derivation the external field is applied along the $+y$ direction, and the canting of the magnetization away from this equilibrium direction is assumed to be small. As a result the magnetization can be described by a (elliptical) precession in the $x,z$ plane. With the inclusion of a $kz$ phase term to allow standing spin waves in the $z$ direction, the magnetization is described by
\begin{equation}
\vec{M} = \begin{bmatrix}
    M_{x} \\[0.3em]
    M_{y} \\[0.3em]
    M_{z}
     \end{bmatrix}
     = \begin{bmatrix}
    M_{x,0} \cos\left( \omega t + k z \right) \\[0.3em]
    M_{\mathrm{s}}                          \\[0.3em]
    M_{z,0} \sin\left( \omega t + k z \right)
     \end{bmatrix}.
\label{Eq:Magnetization}
\end{equation}
The contributions to the effective field include the applied field $\vec{H}_{\mathrm{app}}$, the demagnetization field (originating from shape anisotropy), $\vec{H}_{\mathrm{dem}}$, the anisotropy field (due to the interface anisotropy), $\vec{H}_{\mathrm{ani}}$, and the exchange field $\vec{H}_{\mathrm{ex}}$, and is given by
\begin{equation}
\vec{H}_{\mathrm{eff}} = \begin{bmatrix}
    0 \\[0.3em]
    H_{\mathrm{app}}                          \\[0.3em]
    \left( \frac{2 K_{\mathrm{s}}}{\mu_{0} d M_{\mathrm{s}}^{2}} - 1 \right) M_{z}
     \end{bmatrix}+\frac{2 A_{\mathrm{ex}}}{\mu_{0} M_{\mathrm{s}}^{2}} \nabla^{2} \vec{M}.
\label{Eq:EffectiveField}
\end{equation}
In this equation the thickness of the magnetic layer is given by $d$, and $A_{\mathrm{ex}}$ represents the exchange stiffness. The dispersion of the spin waves is calculated by inserting Eq.\ (\ref{Eq:Magnetization}) and Eq.\ (\ref{Eq:EffectiveField}) into the LLG equation (ignoring damping),
\begin{equation}
\frac{d\vec{M}}{dt} = -\gamma \mu_{0} \left(\vec{M} \times \vec{H}_{\mathrm{eff}} \right).
\label{Eq:LLG}
\end{equation}
Evaluating the $x$ and $z$ components a system of linear homogeneous equations is obtained, for which non-trivial solutions exist when the determinant of the coefficient matrix vanishes,
\begin{multline}
\det \left|
\begin{matrix}
\omega  \\[0.3em]
-\gamma \mu_{0} \left(H_{\mathrm{app}} + C_{2} k^{2} \right)
\end{matrix}
\right. \\ \left.
\begin{matrix}
\gamma \mu_{0} \left(H_{\mathrm{app}} + M_{\mathrm{s}} + C_{2} k^{2} - C_{1} \right) \\[0.3em]
-\omega
\end{matrix}
\right| = 0
\label{Eq:Determinant}
\end{multline}
in which the following abbreviations are used,
\begin{equation}
C_{1} = \frac{2 K_{\mathrm{s}}}{\mu_{0} d M_{\mathrm{s}}},
\end{equation}
\begin{equation}
C_{2} = \frac{2 A_{\mathrm{ex}}}{\mu_{0} M_{\mathrm{s}}}.
\end{equation}
Solving Eq.\ (\ref{Eq:Determinant}) for $\omega \ (= 2 \pi f)$ leads to the dispersion relation given by
\begin{multline}
f\left(k\right) = \frac{\gamma}{2 \pi} \left[\left(B_{\mathrm{app}} + \frac{2 A_{\mathrm{ex}}}{M_{\mathrm{s}}} k^{2}\right) \right. \\ \left.\left(B_{\mathrm{app}} + \mu_{0} M_{\mathrm{s}} - \frac{2 K_{\mathrm{s}}}{t M_{\mathrm{s}}} + \frac{2 A_{\mathrm{ex}}}{M_{\mathrm{s}}} k^{2}\right)\right]^{\frac{1}{2}}.
\label{Eq:DispersionAex}
\end{multline}
The relation between the exchange stiffness and the spin wave stiffness of long wave-length spin waves is given by\cite{Coey2010}
\begin{equation}
A_{\mathrm{ex}} = \frac{M_{\mathrm{s}} D_{\mathrm{sw}}}{2 \gamma \hbar}.
\end{equation}
Using this relation the dispersion relation of Eq.\ (\ref{Eq:DispersionAex}) can be rewritten to
\begin{multline}
f\left(k\right) = \frac{\gamma}{2 \pi} \left[\left(B_{\mathrm{app}} + \frac{D_{\mathrm{sw}}}{\gamma \hbar} k^{2}\right) \right. \\ \left.\left(B_{\mathrm{app}} + \mu_{0} M_{\mathrm{s}} - \frac{2 K_{\mathrm{s}}}{t M_{\mathrm{s}}} + \frac{D_{\mathrm{sw}}}{\gamma \hbar} k^{2}\right)\right]^{\frac{1}{2}}.
\end{multline}

\section{}
\label{app:CalculationLinMs}

In this section the calculation of the efficiency as a function of the Co thickness for the non-collinear bilayer with N = 4 and a wedge shaped top Co layer is shown again, now including a thickness dependent saturation magnetization of the Co layer. The reason for this alternative calculation is shown in Fig.\ \ref{Fig:SCAbsorption}(a) of the main article. In this figure it was seen that the measured dispersion of the fundamental precession in the wedged Co layer (filled dots) is not well described by the Kittel relation (solid lines). For all field strengths it seems that there is an additional thickness dependence that is not captured by Eq.\ (\ref{Eq:KittelIP}). As discussed in the main article, one possibility is that the saturation magnetization of the Co layer decreases with decreasing layer thickness, as a result of a lower exchange constant in the interface regions due to interface intermixing. In the following a first approximation of the dependency of the saturation magnetization on the layer thickness is given, assuming a constant surface anisotropy. Afterwards, the efficiency as a function of the top Co layer thickness is calculated again, including the found relation between the saturation magnetization and Co layer thickness. It will be shown that the efficiency as a function of the Co thickness for this alternative analysis is similar as was found in the main article using a constant $M_{\mathrm{s}}$.

\begin{figure}
\includegraphics[scale=0.3,trim = 0mm 0mm 0mm 0mm,clip]{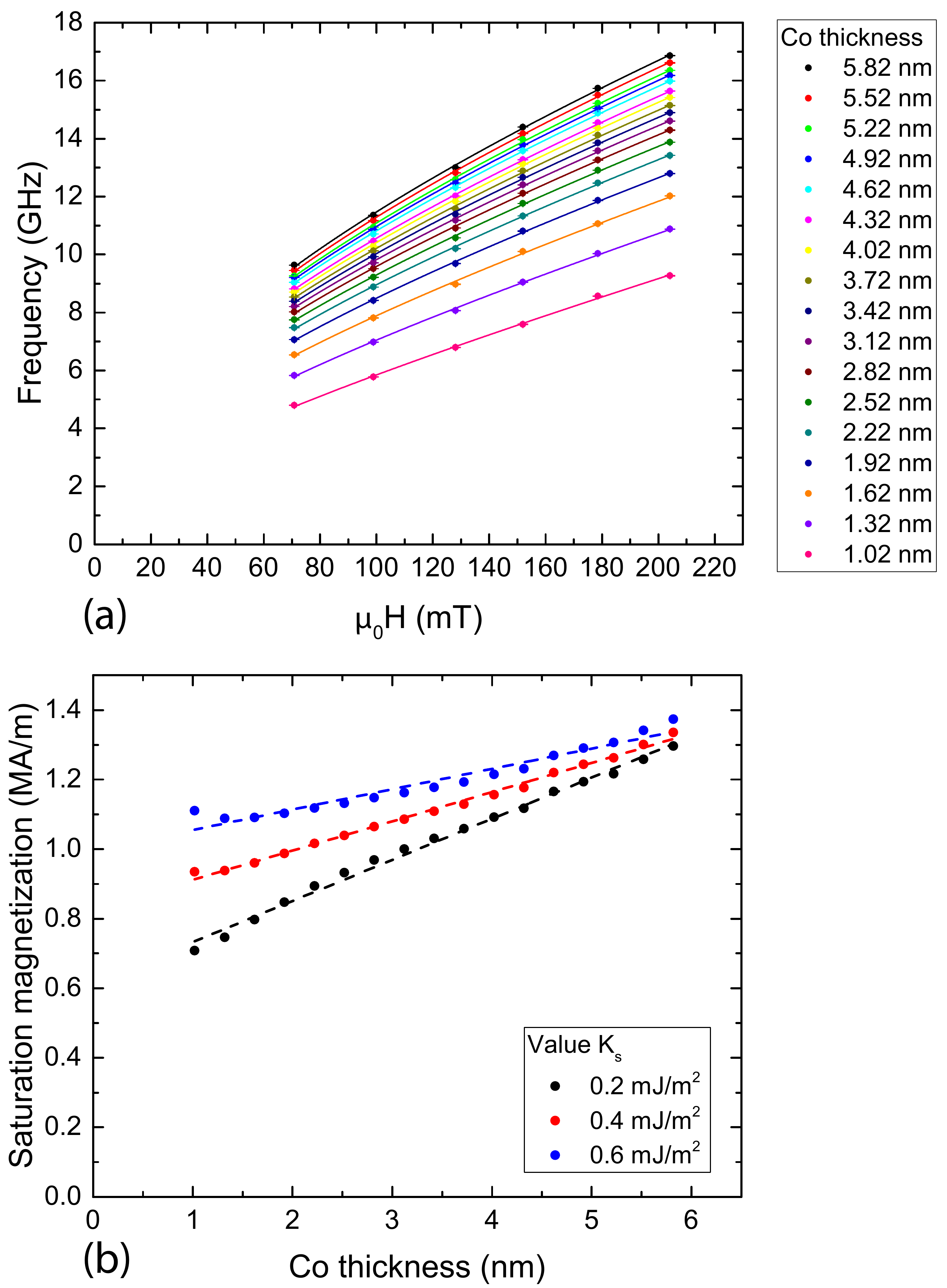}
\caption{(a) Precession frequency measured as a function of top Co layer thickness for six different in-plane applied fields. The measurements are performed on a non-collinear magnetic bilayer with $N = 4$ and wedged top Co layer with thickness ranging from $0$ nm to $6$ nm. The solid curves represent fits to the data using the Kittel relation using an effective magnetization to include both the saturation magnetization and surface anisotropy. (b) Calculated saturation magnetization as a function of Co layer thickness for three different values for the surface anisotropy. The dashed lines show that the relation between the saturation magnetization and Co thickness is well described by a linear dependency for each value of the surface anisotropy.}
\label{Fig:KittelFits&LinearMs}
\end{figure}

To find the relation between the Co thickness and the saturation magnetization, the data of Fig.\ \ref{Fig:SCAbsorption}(a) is reevaluated in Fig.\ \ref{Fig:KittelFits&LinearMs}(a) (filled dots). In this figure the precession frequency is plotted as a function of the applied field, for different thicknesses of the Co layer. The data for each thickness is fitted using a slightly different version of the Kittel relation (solid curves),
\begin{equation}
    f_{\mathrm{IP}} = \frac{\gamma}{2 \pi} \sqrt{B_{\mathrm{app}} \left( B_{\mathrm{app}} + \mu_{0} M_{\mathrm{eff}} \right)},
\label{Eq:KittelIPMeff}
\end{equation}
introducing the effective magnetization ($M_{\mathrm{eff}}$) in order to get the best fit and not yet make any assumption about both $M_{\mathrm{s}}$ and $K_{\mathrm{s}}$. The effective magnetization is given by
\begin{equation}
    M_{\mathrm{eff}} = M_{\mathrm{s}} - \frac{2 K_{\mathrm{s}}}{\mu_{0} M_{\mathrm{s}} t}.
\label{Eq:Meff}
\end{equation}
It can be seen that substitution of Eq.\ (\ref{Eq:Meff}) into Eq.\ (\ref{Eq:KittelIPMeff}) returns the Kittel relation as given in Eq.\ (\ref{Eq:KittelIP}). With $M_{\mathrm{eff}}$ determined for each thickness, the saturation magnetization can be calculated when the surface anisotropy at each thickness is known. Since a wedge is used, the adjacent layers of the Co layer are identical for each thickness. Therefore, in this first approximation it is assumed that the surface anisotropy is constant throughout the wedge, and thus the same for each Co thickness. The exact value of $K_{\mathrm{s}}$ is not known, and can not be determined from the data in Fig.\ \ref{Fig:KittelFits&LinearMs}(a). Therefore, the saturation magnetization is calculated using a range of different values for $K_{\mathrm{s}}$, ranging between $0.2 - 0.6$ mJ.m$^{-2}$ derived from literature \cite{Johnson1996}. The calculated saturation magnetization as a function of Co layer thickness for three different values for $K_{\mathrm{s}}$ are shown in Fig.\ \ref{Fig:KittelFits&LinearMs}(b) (solid dots). It can be seen that the relation between the saturation magnetization and Co layer thickness is well approximated by a linear dependency for all three values of $K_{\mathrm{s}}$. Therefore, as a first estimation a linear thickness dependency of $M_{\mathrm{s}}$, given by $M_{\mathrm{s}}(t_{\mathrm{Co}}) = M_{\mathrm{s},0} + A \ t_{\mathrm{Co}}$, is added to the Kittel relation. Using this extended Kittel relation the data presented in Fig.\ \ref{Fig:SCAbsorption}(a) is fitted again, now using $A$, $M_{\mathrm{s},0}$ and $K_{\mathrm{s}}$ as (global) fitting parameters. As can be seen in the figure, the fits including this thickness dependent saturation magnetization describe the measured data with much more accuracy (dashed curves).

\begin{figure}
\includegraphics[scale=0.3,trim = 0mm 0mm 0mm 0mm,clip]{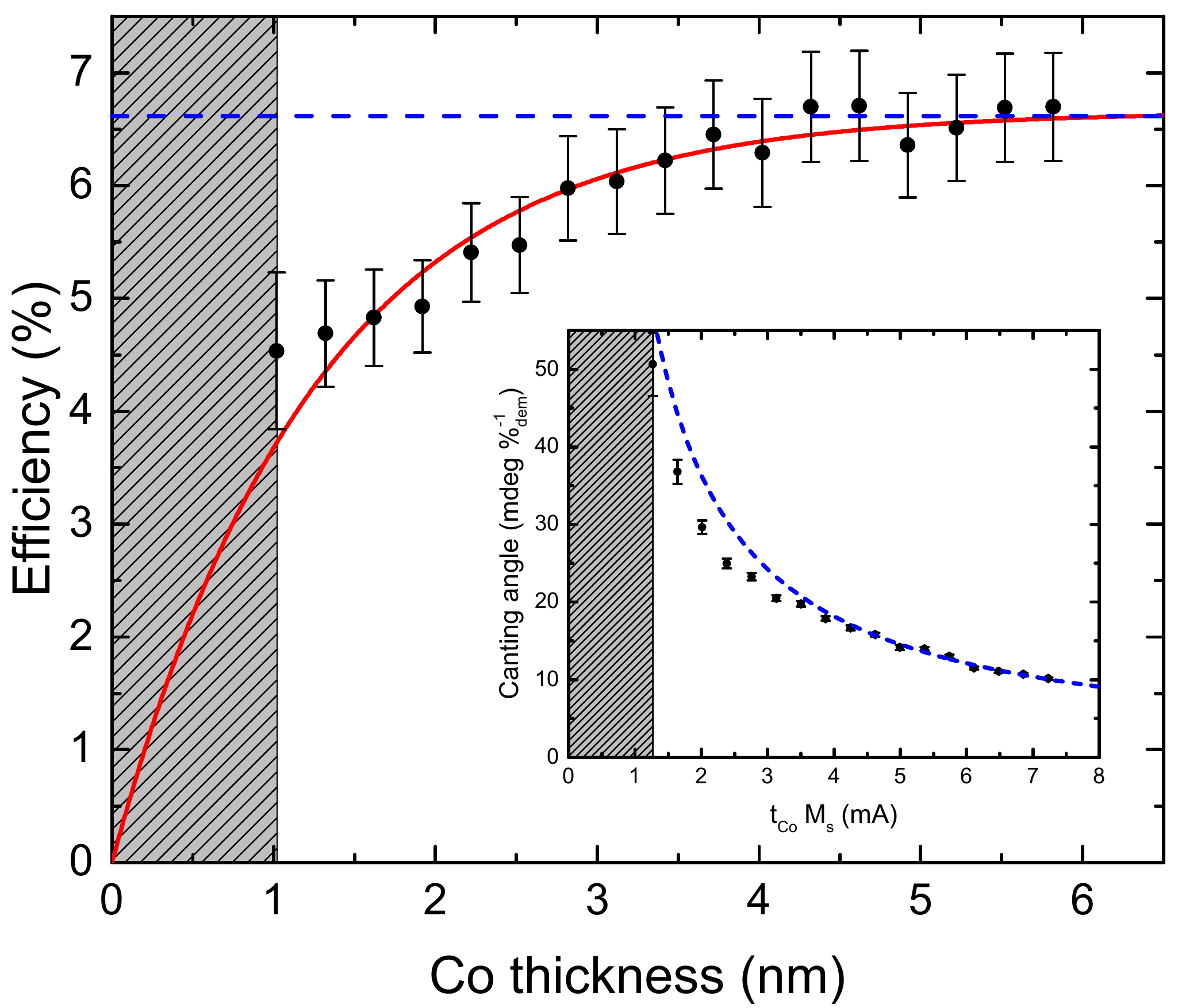}
\caption{Efficiency and canting angle per percent demagnetization (inset) as a function of top Co layer thickness, determined using a linear dependency of the saturation magnetization on the Co layer thickness. The solid curve represents a fit to the data. The dotted lines represent constant efficiency and corresponding $\theta_{\mathrm{c,\%}}$, which describes the case when there is full absorption independent of top Co layer thickness.}
\label{Fig:SCAbsorptionLinearMs}
\end{figure}

Using the values obtained for $A$, $M_{\mathrm{s},0}$ and $K_{\mathrm{s}}$, the efficiency as a function of the Co thickness is calculated again, and presented in Fig.\ \ref{Fig:SCAbsorptionLinearMs}. It can be seen that the overall (exponential) behaviour is similar as was found in the main article using a constant $M_{\mathrm{s}}$ and $K_{\mathrm{s}}$ throughout the wedge. However, there are some small differences. In the present case the exponential decrease of the efficiency for small Co thicknesses is more pronounced. Also, the characteristic absorption depth is larger, $\lambda_{\mathrm{Co}} = 1.24 \pm 0.08$ nm compared to $\lambda_{\mathrm{Co}} = 0.96 \pm 0.07$ nm for constant $M_{\mathrm{s}}$ and $K_{\mathrm{s}}$. This results in $90\%$ being absorbed within the first $2.9 \pm 0.2$ nm compared to the $2.2 \pm 0.2$ nm in the main article. Lastly, the saturation value of the efficiency is a bit smaller, $\epsilon_{\mathrm{max}} = 6.7 \pm 0.1$ \% compared to $\epsilon_{\mathrm{max}} = 7.3 \pm 0.1$ \% in Fig.\ \ref{Fig:SCAbsorption}(b).

In conclusion, it has been demonstrated that the overall behaviour of the efficiency as a function of the Co thickness is similar in both cases, i.e.\ with constant $M_{\mathrm{s}}$, or with a thickness dependent $M_{\mathrm{s}}$. This shows that the conclusion on the limited absorption depth is robust. Including the thickness dependence of the saturation magnetization to the Kittel relation greatly improved the fits to the data, indicating that the thickness dependent saturation magnetization might be a viable approximation.

\end{document}